\documentclass[12pt]{article}

\usepackage{aas_macros}

\usepackage{geometry}                
\usepackage{fullpage}  

\usepackage{fourier} 
\usepackage[scaled=0.875]{helvet} 

\usepackage[OT1,T1]{fontenc}

\usepackage[colorlinks=true, linkcolor=blue]{hyperref}

\usepackage[
natbib=true,     		
style=authoryear,    	
backref=true               
]{biblatex}  

\usepackage{graphicx}

\title{Turbulent magnetic field
	amplification driven by cosmic-ray pressure gradients}
	
\author{L. O'C. Drury$^{1}$ and T. P. Downes$^{1,2,3}$ \\
\url{mailto:ld@cp.dias.ie}\\
\url{mailto:turlough.downes@dcu.ie}\\[2ex]
$^{1}$School of Cosmic Physics, Dublin Institute for Advanced Studies,\\ 31 
Fitzwilliam Place, Dublin 2, Ireland \\[1ex]
$^{2}$School of Mathematical Sciences, Dublin City University,\\
	      Glasnevin, Dublin 9, Ireland\\[1 ex]
$^{3}$National Centre for Plasma Science and Technology,\\ Dublin City 
University, Glasnevin, Dublin 9, Ireland}

\begin{document}
\maketitle

\sloppy

\begin{abstract}
Observations of non-thermal emission from several supernova remnants suggest 
that magnetic fields close to the blastwave are much stronger than would be 
naively expected from simple shock compression of the field permeating the 
interstellar medium (ISM).

We present a simple model which is capable of achieving sufficient magnetic 
field amplification to explain the observations.  We propose that
the cosmic-ray pressure gradient acting on the inhomogeneous ISM 
upstream of the supernova blastwave induces strong turbulence upstream
of the supernova blastwave.  The turbulence is generated through the 
differential acceleration of the upstream ISM which occurs as a result of 
density inhomogeneities in the ISM.  This turbulence then amplifies the 
pre-existing magnetic field.

Numerical simulations are presented which demonstrate that amplification
factors of 20 or more are easily achievable by this mechanism when
reasonable parameters for the ISM and supernova blastwave are assumed.
The length scale over which this amplification occurs is that of the
diffusion length of the highest energy non-thermal particles.
\end{abstract}

\section{Introduction}

The idea that magnetic fields might be substantially amplified by
cosmic-ray driven processes in strong shocks, and in particular those
bounding young supernova remnants (SNRs), has recently attracted
considerable attention.  Much of this derives from the seminal work of
\citet{2004MNRAS.353..550B} who pointed out the existence of a strong
current-driven instability under conditions thought to be appropriate to
young remnants.  It is also supported by a range of indirect, but quite
compelling, observational arguments which point to substantially higher
effective magnetic fields at the shocks of young remnants than would be
expected just from adiabatic compression of a pre-existing interstellar
field \citep{2003ApJ...584..758V, 2003A&A...412L..11B, 2004ApJ...602..257B, 2005ApJ...621..793B, 2005A&A...433..229V, 2006AdSpR..37.1902B,2006A&A...453..387P,2012A&ARv..20...49V,Uchiyama:2007ly,2004A&A...416..595Y}.  The idea is very attractive because it appears to fill in a missing piece in the theory of cosmic ray acceleration by SNR shocks and allow acceleration to the energies needed to explain the cosmic-ray `knee' particles; if the magnetic field strength is only a few $\mu$G as expected in the interstellar medium it is very hard to get the acceleration to reach these energies as pointed out by \citet{1983A&A...125..249L}. 

 Actually the idea has a long, if not widely known, pre-history.  Over half a century ago \citet{1960MNRAS.120..338H}  speculated that collisionless interstellar shocks could dissipate kinetic energy into either thermal energy, cosmic-rays or magnetic field energy and over a quarter of a century ago \citet{Cowsik:1980qy} pointed out the need for substantially amplified fields in Cas-A on the basis of early gamma-ray observations.

The Bell mechanism, which essentially relies on the current carried by non-magnetised high-energy cosmic-ray particles driving a return current in the thermal plasma, definitely can occur in the precursor region of SNR shocks if these are strong particle accelerators.  However it need not be the only process and indeed it suffers from the disadvantage that it can only generate fields on scales smaller than the gyro-radius of the driving particles.  Without some inverse cascade or other process these fields are thus on scales too small to be used to accelerate the highest energy particles themselves.  It is thus of interest to examine other possible mechanisms.  As pointed out \citep{Diamond:2007yq,Malkov:2009kx} one promising candidate is the instability identified in \citet{Drury:1986uq} and further studied in \citet{Begelman:1994fj}; {\it cf} also \citet{Webb:1999vn} and \citet{Ryu:1993rt}  This can have faster growth rates than the Bell instability and has the great advantage of operating on scales large compared to the gyro-radius of the driving particles and in relying only on rather simple and robust physics.


\citet{2009ApJ...707.1541B} proposed an analytic model, similar to that investigated here, in which the cosmic ray pressure drives
small-scale dynamo action which amplifies the pre-shock magnetic field.  The magnetic field is amplified in the usual stretch-fold
manner by the solenoidal component of the velocity field in the precursor.  This solenoidal component of the velocity field
results from the inhomogeneous density of the medium upstream of the shock (see section \ref{sec:model}).  These authors find that significant 
amplification of the magnetic field is possible through this mechanism.  In this work, we focus on using fully nonlinear numerical 
simulations in which both solenoidal and compressive components of the precursor velocity field contribute to the process of 
amplification of the magnetic field.  These two approaches to the problem can be seen as complementary, with each having its own 
strengths.

\section{Physical basis for the instability and toy model}
\label{sec:model}

The instability arises very simply and generally from the fact that the cosmic ray pressure gradient in the shock precursor exerts a local ponderomotive force on the thermal plasma which will not in general be proportional to the mass density.  Density fluctuations thus induce acceleration fluctuations, which lead in turn to velocity fluctuations which then induce further density fluctuations.  In the case of linear perturbations of an essentially homogeneous isentropic background plasma these fluctuations are acoustic (or magneto-acoustic) modes and the process leads to the acoustic instability discussed in \citet{Drury:1986uq}, but more generally one should also consider entropy fluctuations.   In one dimension the instability can be suppressed if the diffusion coefficient for the cosmic-rays is rather artificially chosen to be inversely proportional to the mass density, but it is impossible to suppress the instability in more than one dimension.  If the distribution of cosmic-ray pressure is adjusted to avoid instability perpendicular to the shock front, it is unstable parallel and vice-versa.  In general the scattering experienced by the cosmic rays, and thus the effective diffusion coefficient,  is a complicated function of the local magnetic field strength and power-spectrum of magnetic irregularities.  It will thus change if the plasma is locally adiabatically compressed, and this will feed back into the cosmic-ray distribution and thus the cosmic-ray pressure gradients, but in a non-obvious way.   Rather than try to model this we consider the simplest possible case where the cosmic-ray propagation is totally decoupled from the matter dynamics.  

This is something of an extreme assumption and thus requires some discussion.  It amounts to assuming that the diffusion
tensor of the cosmic ray particles is a given function of position and momentum, and is not affected by the magnetic field and
density fluctuations.  Now clearly at one level this is nonsense.  The magnitude of the diffusion and the anisotropy of the
tensor are intimately related to the local magnetic field strength and orientation, and the strength of the scattering wave field
will also be influenced by the local compression or expansion of the medium in which the waves are propagating.  But these
effects are very complicated and beyond our current ability to model in any realistic way.  To simply set them to zero is, while
unnatural, not unphysical, and leads to a very simple toy model.  The key question is whether, in doing this, we have thrown out
any essential physics and we do not think that we have done so.  On the contrary, as all analyses of the acoustic instability
show, it is very difficult to suppress the instability (and in fact impossible in more than one spatial dimension).  If
anything, easier diffusion of cosmic ray particles along channels evacuated by increased cosmic ray pressure should increase the
instability aided by an analogue of the Parker instability in the shock precursor (cosmic ray inflated magnetic loops will experience an effective buoyancy in the apparent gravitational field resulting from the precursor deceleration).  In this
situation to simply switch off these complications and treat a simple toy model in which the essential features of the
instability are retained seems a sensible thing to do and we do not think that it either artificially enhances or suppresses the
importance of the instability.  If a simulation with a more realistic coupling between the cosmic ray
pressure and the dynamical perturbations can be carried out we will be delighted, but we will be surprised if the results are radically different.

Motivated by \citet{Malkov:1999ys} and his universal asymptotic solution for strong accelerators, which has a linearly rising
cosmic ray pressure in the precursor, we thus consider a toy system consisting of a rectangular computational box extending in
the $x$-direction from $0$ to $L$ within which the cosmic ray pressure $P_{\rm C}$ rises linearly from zero at the inflow side to a
value of order the ram pressure of the inflowing plasma at the outflow side.  The shock position is thus taken to be at
$x=L$.
\begin{equation}
P_{\rm C}(x) =  \theta \rho_0 U_0^2 {x\over L},
	\label{eqn:pc_def}
\end{equation}
where $0<\theta<1$ is a positive parameter less than unity.

The flow is thus decelerated by a uniform body force $-\theta\rho_0U_0^2/L$ representing the reaction of the accelerated cosmic
rays (and the work done is of course the work done in accelerating them).  We then seed the inflowing plasma, which is treated
as an ideal MHD fluid, with small-scale density fluctuations and follow the evolution of the resulting turbulence and magnetic
field amplification.  Because the computational box is intended to cover the shock precursor region with a shock sitting just
downstream on the high-$x$ side of the right-hand boundary at $x=L$ the boundary conditions are pure inflow on the left and pure outflow on the right.  It is necessary to choose $\theta$ such that this condition is satisfied and no characteristic curves re-enter the computational domain from downstream (on the right).

This model has the great advantage that it captures the essential physics of the instability, a bulk force acting on the plasma which is not proportional to the local density, without having to compute the cosmic ray pressure distribution and thereby reduces the problem to a pure computational MHD one. 

If we assume that the incoming flow contains density irregularities of magnitude $\delta \rho$ on a length scale $\lambda$ the bulk force, operating on a time scale of order the advection time through the precursor, will generate velocity fluctuations of magnitude
\begin{equation}
\delta u \approx {\delta \rho\over \rho_0} {1\over\rho_0} {\theta \rho_0^2 U_0\over L} {L\over U_0} \approx {\delta\rho\over\rho_0} \theta U_0
\label{eqn:delta-u}
\end{equation}
on the same length scale $\lambda$.  If this is to drive turbulence we require the eddy turn-over time to be short compared to the outer-scale and thus
\begin{equation}
{\lambda\over\delta u} \ll {L\over U_0} \Rightarrow \lambda \ll \theta {\delta\rho\over\rho_0} L
\label{eqn:lambda-limit}
\end{equation}
Density fluctuations satisfying this not very restrictive condition should be capable of inducing turbulence and thus magnetic field amplification.    The total amount of kinetic energy available in the turbulence can be roughly estimated as 


\begin{equation}
\label{eqn:e_f}
e_{\rm F} = {1\over 2}\rho_0 (\delta u)^2 \approx {1\over 2 \rho_0 }\left(\delta\rho\right)^2 \theta^2 U_0^2
\end{equation}
and thus the maximum amplified field should be below full equipartition by a 
factor of order $\theta^2 (\delta\rho/\rho_0)^2$.  If nonlinear effects drive the density fluctuations to saturation at $\delta\rho\approx \rho$ (as is probable) then this process could be very efficient at converting flow energy into magnetic energy if $\theta \approx 1$.

That turbulence can amplify magnetic fields at the blastwaves of 
supernova shock remnants has been proposed previously
\citep[e.g.][]{2007ApJ...663L..41G, 2012ApJ...747...98G}.  In these works the 
field amplification occurs downstream of the shocks and the turbulence is 
driven by vorticity created as an inhomogeneous fluid passes through a strong 
shock.  Our model is quite different in that field amplification occurs 
in the upstream medium and the turbulence is driven by the cosmic ray 
pressure.  It is reasonable to expect that the process examined by
\citet{2007ApJ...663L..41G} and \citet{2012ApJ...747...98G} will then
operate on the cosmic-ray amplified field to further amplify it in the
downstream region.

\section{Numerical method}

To investigate this model further we employ numerical simulations of a
decelerating, ideal magnetohydrodynamic (MHD) flow using the HYDRA code \citep{OSullivan:2006zr,OSullivan:2007mz}
set to simulate ideal MHD, rather than multifluid
MHD.  The equations solved are
\begin{eqnarray}
\frac{\partial \rho}{\partial t} + \mathbf{\nabla} \cdot \left(\rho
		      \mathbf{u}\right)  & = & 0 \label{mass} \\
\frac{\partial \rho \mathbf{u}}{\partial t} + \nabla\cdot\left( \rho \mathbf{u} 
		\mathbf{u} + P \mathbf{I}\right) & = & \mathbf{J}\times\mathbf{B}
+ \mathbf{F}_{\rm cr}, \label{neutral_mom} \\
\frac{\partial e}{\partial t} + \mathbf{\nabla} \cdot \left[ \left(e +
		P\right)\mathbf{u}\right] & = &
\mathbf{J}\cdot(\mathbf{u}\times\mathbf{B}) + \mathbf{F}_{\rm
	cr}\cdot\mathbf{u}\\
\frac{\partial \mathbf{B}}{\partial t} +
\nabla\cdot(\mathbf{u}\mathbf{B}-\mathbf{B}\mathbf{u}) & = &0 , \label{B_eqn} \\
								  \nabla\cdot\mathbf{B} & = & 0 \label{divB}
\end{eqnarray}
where $\rho$ is the mass density, $\mathbf{u}$ is the fluid velocity,
$P$ is the thermal pressure, $\mathbf{B}$ is the magnetic field,
$\mathbf{F}_{\rm cr}$ is the force due to the cosmic ray pressure
gradient and $\mathbf{I}$ is the identity matrix.  $\mathbf{F}_{\rm cr}$
is given by
\begin{eqnarray}
\mathbf{F}_{\rm cr} & = & - \mathbf{\nabla} P_{\rm C} \nonumber \\
		& = & - \frac{\theta \rho_0 U_0^2}{L}\mathbf{\hat{\i}} 
\end{eqnarray}
\noindent (see equation \ref{eqn:pc_def}).

These equations are advanced in time using a standard van Leer-type second order, finite volume, shock
capturing scheme.  The magnetic field divergence is controlled using the
method of \citet{Dedner:2002fr}.  This slightly unusual form of the MHD
equations is used as HYDRA is a multifluid code, making
this form of the equations more convenient.  This code has been
extensively validated for both multifluid and ideal MHD set-ups.

For the simulations presented in this work we take $\theta$ (defined in
equation \ref{eqn:pc_def}) to be 0.6 which is observationally reasonable
\citep[e.g.][and references therein]{2012A&ARv..20...49V}.  For the large Mach 
numbers associated with supernova blastwaves this will give us a very
significant acceleration of the pre-shock flow.

\subsection{Definition of the problem}

\label{sec:problem-def}

We wish to simulate a flow which is being accelerated by a constant
force in front of a supernova blastwave.  The constant force is exerted
on the fluid through interactions with a non-thermal particle
population.  The density of the pre-shock fluid is taken to be inhomogeneous 
and, for consistency with expectations for isothermal turbulence in the 
interstellar medium, it is taken to have a log-normal probability density 
function.

We formulate our problem in the rest frame of the blastwave which is
assumed to be planar and to have a Mach number of 100.  The
position of the blastwave is taken to have a value of $x = L$.  The 
computational domain is given length $L=1$ in the $x$ direction, and length 
$\frac{L}{8}$ in the $y$ direction.  The length in the $z$ direction is one 
grid zone for our 2D simulations.  This 
means that the blastwave itself is at the boundary of our domain.  This is 
appropriate as our focus in this work is on the development of the instability 
in the pre-shock fluid.

A further advantage of this set-up is that, for appropriate definition
of $\theta$ (see Equation \ref{eqn:pc_def}), the conditions at $L=1$
should ensure that no information propagates from outside the domain at
this point as the flow will remain supersonic in the positive $x$
direction across the entire grid.  There are some subtleties to this,
however, and this is discussed further in Section
\ref{sec:boundary-conditions}.

A crucial parameter for determining whether or not significant magnetic 
field amplification can occur is the ratio, in the ``mean'' rest frame of the 
fluid, of the energy residing in the magnetic field to that residing in the 
motions resulting from the stochastic density distribution and the 
(mass-independent) body force exerted by the cosmic rays.

Equation \ref{eqn:e_f} gives the kinetic energy associated with the expected 
velocity fluctuations arising from the differential acceleration due to the 
cosmic ray pressure.  In order for these fluctuations to amplify the initial 
magnetic field without significant back-reaction from the magnetic field on 
the fluid motions, we require that the energy density in the initial field be 
much less than that associated with the fluctuations:
\begin{equation}
e_{\rm F} >> e_B
\end{equation}
where $e_B$ is the energy density associated with the magnetic field.  Thus 
we require
\begin{equation}
\frac{B_0^2}{2} << \frac{\left(\delta \rho \theta U_0\right)^2}{2 \rho_0}
\end{equation}
or
\begin{equation}
B_0 << \frac{\delta \rho \theta U_0}{\sqrt{\rho_0}}
\label{eqn:b_limit}
\end{equation}
For supernova blastwaves propagating into the ISM we certainly expect
this condition to be satisfied.

\subsection{Initial conditions}

The initial distribution of density is defined in the following way.
First we define a function $f$ by
\begin{equation}
f(x,y,z) = \sum_{k_x,k_y,k_z}
A_i \sin\left\{\frac{2 \pi}{L}\left[k_x x +
		 k_y y + k_z z \right] + \phi_i\right\}
\end{equation}
where $i$ is an index for the wave-vector.  $A_i$ and $\phi_i$ are
random variables picked from the ranges $[0,1]$ and $[0, 2 \pi]$,
respectively and $\alpha$ is a normalising constant to give the desired
RMS of the density distribution.  The wavenumbers, $k_x$, $k_y$ and
$k_z$ range from -32 to 32, -4 to 4 and -4 to 4, respectively.  The density 
distribution itself is then given by
\begin{equation}
\rho(x,y,z) = e^{\alpha f(x,y,z)}
\label{eqn:rho-def}
\end{equation}
For all simulations presented in this work this RMS variation is chosen to be 
0.2, while the average density is approximately 1.  This recipe gives us a 
log-normal distribution for the density distribution which is typical of what 
would be expected in the case of pre-existing (isothermal) turbulence.

The fluid initially has a uniform pressure, chosen to give a mean sound 
speed of 1, and it is taken to have a ratio of specific heats of $5/3$.  We 
perform the simulations in a frame of reference taken to be the rest frame of 
the supernova blastwave (see Section \ref{sec:problem-def}) and give the fluid 
a uniform speed, $U_0$, of 100 in the positive $x$ direction.  Equation
\ref{eqn:lambda-limit} then suggests that we require the scale of our
density inhomogeneities to satisfy
\begin{equation}
\lambda << 0.12
\end{equation}
which, as can be seen from Equation \ref{eqn:rho-def}, is satisfied (at
least marginally) for the higher values of $k_x, k_y$ and $k_z$.

The magnetic field is also initially uniform.  The field is
chosen to be purely in the $y$ direction and so these simulations are
appropriate for a perpendicular shock.  Equation \ref{eqn:b_limit}
is clearly satisfied for the values of $|\mathbf{B}_0|$ and $(\delta
\rho)_{\rm rms}$ given in Table \ref{table:nomenclature}.  Thus we do expect 
to get significant magnetic field amplification, at least until the field is 
amplified to levels at which the magnetic energy density becomes of order the 
kinetic energy density of the fluctuations.

Table \ref{table:nomenclature} contains a summary of the simulations run.  We 
first perform a resolution study with $\delta \rho_{\rm rms} = 0.2$ and 
$B_0 = 0.1$ to investigate the convergence of our simulations.  We also 
briefly investigate the influence of varying the initial magnetic field 
strength and the value of $\delta \rho_{\rm rms}$.

\begin{table}
\caption{Summary of the simulations used in this work. \label{table:nomenclature}}
\begin{tabular}{lcccc}\hline
Simulation & Grid size & $\delta \rho_{\rm rms}$ & $B_0$ & Plasma
$\beta$ \\ \hline
1 & $500 \times 64$ & 0.2 & 0.1 & 200 \\
2 & $1000 \times 125$ & 0.2 & 0.1 & 200 \\
3 & $2000 \times 250$ & 0.2 & 0.1 & 200 \\
4 & $4000 \times 500$ & 0.2 & 0.1 & 200 \\
5 & $4000 \times 500$ & 0.44 & 0.1 & 200 \\
6 & $4000 \times 500$ & 0.2 & 0.01 & 2000  \\
\end{tabular}
\end{table}

\subsection{Boundary conditions}
\label{sec:boundary-conditions}

All of the boundary conditions are set to periodic with the exception of the 
$yz$-planes at $x = 0$ and $x = L$.  At $x=0$ the speed, pressure and
magnetic field of the fluid are fixed at $U_0$, 0.6 and $|\mathbf{B_0}|$ 
respectively.  The density condition varies with time and is given by 
\begin{equation}
\rho(t,y,z) = \exp\left\{\alpha f(- k_x U_0 t, y, z)\right\}
\end{equation}
so that the density distribution flowing onto the computational domain
at $x=0$ is that of the initial distribution of the density.  This gives
the overall simulation a periodicity of 0.01, the flow time across the
grid in the absence of the cosmic ray pressure.

The boundary conditions at $x=L$ are set to gradient zero, with the
exception of the pressure which is fixed at 0.6.  The pressure is fixed
in this way as gradient zero boundary conditions for systems such as this, 
where there is supersonic flow out of the computational domain but with varying
pressure, can occasionally lead to spurious high pressure waves being
driven into the domain from the boundary.

\subsection{Determining the magnetic field amplification}

In order to determine the magnetic field amplification for a given
simulation we proceed as follows.  We analyse snapshots of the
simulation taken at intervals of $\Delta t = 0.002$ between time
$t=0.02$ and $t=0.04$ by averaging the magnetic field strength over $0
\leq y \leq \frac{L}{8}$ for each value of $x$ to give us
$<B(x,y,t)>_{y\in \left( 0,\frac{L}{8}\right)}$.
This quantity is then averaged over all snapshots taken between $t=0.02$
and $t=0.04$ to give us a time-averaged magnetic field strength as a
function of $x$: $<<B(x,y,t)>_{y\in \left( 0,\frac{L}{8}\right)}>_{t \in
	(0.02,0.04)}$.  This is then normalised by the initial
field strength in order to determine the magnetic field amplification as
a function of $x$.

\section{Resolution study}
\label{sec:res-study}

\begin{figure}
\includegraphics[width=9cm]{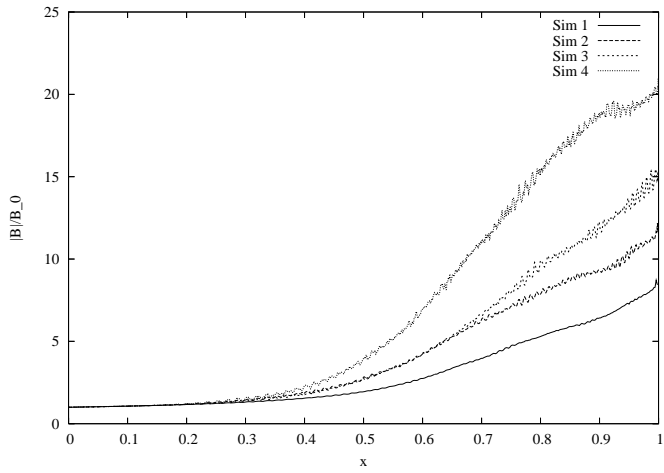}
\caption{ \label{fig:res-study} Plots of the magnetic field
amplification, averaged over $0 \leq y \leq \frac{L}{8}$, for the
simulations in the resolution study.}
\end{figure}

The amplification of the magnetic field is plotted as a function of $x$
in figure \ref{fig:res-study} for simulations 1 -- 4 (defined in Table 
\ref{table:nomenclature}).  Interestingly, the magnetic field
amplification does not appear to be converged, even at the relatively
high resolution of Simulation 4.  This deserves some consideration.

In this system, the turbulence generated is driven by the cosmic ray
pressure acting on parcels of fluid with differing densities.  Thus the
length scales on which the turbulence is driven are those on which the
density varies.  As can be seen by inspection of figure
\ref{fig:rho-dist}, these length scales range from almost the full
range of $y$ right down to the shortest length scales resolved by the
simulation.  This will always be the case unless the simulation resolves
length scales down to the physical dissipation scale (determined either
by viscous effects or non-ideal MHD effects).  This is quite different
to the problem of modelling general turbulence in the ISM where the
driving scales are taken to be large, and the energy then cascades down
to the dissipation length scale.  In this case one can hope to simulate
at least part of the ``inertial range'', but in the system being
modelled here there is inherently no inertial range.

However, things are not quite as bad as they seem.  It is clear that
higher resolution gives us greater magnetic field amplification.  This
is what we would expect as, with higher resolution, we are allowing the
field to be amplified on a greater range of length scales, and hence we
might expect to get higher overall amplification.  We expect, then, that
as we increase our resolution we will continue to get more magnetic
field amplification until either $e_{\rm F} \approx e_B$ or the
simulation resolves the relevant dissipation scale of the system.

Thus the results of the resolution study imply that the magnetic field
amplification levels presented in this work are, in fact, lower limits
for what would actually happen in the ISM immediately upstream of a
supernova blastwave.

\section{Results}

\begin{figure*}
\includegraphics[width=15cm]{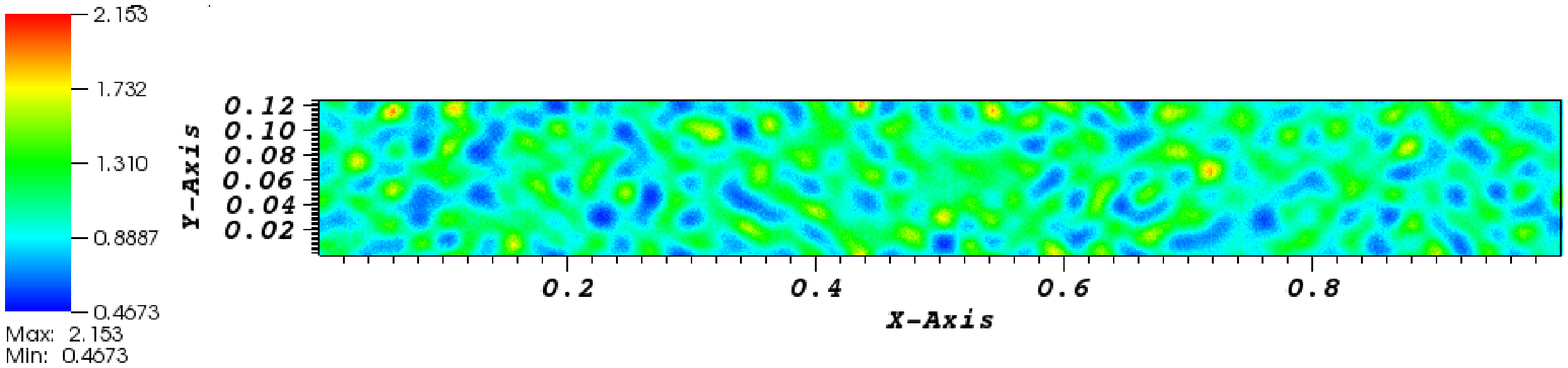}
\includegraphics[width=15cm]{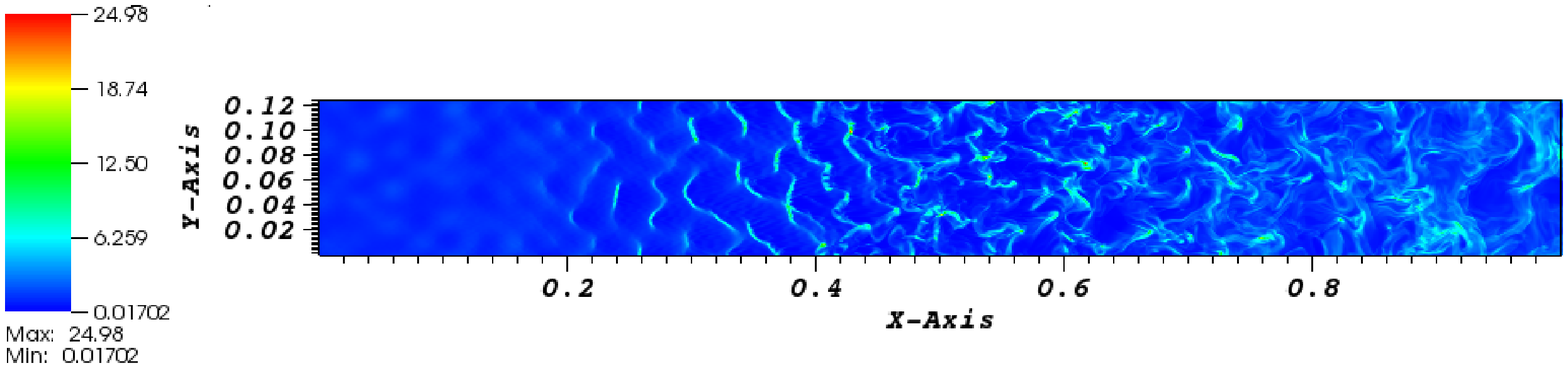}
\caption{\label{fig:rho-dist} Plots of the distribution of the density at $t=0$ (top panel)
	and $t=0.04$ (bottom panel).}
\end{figure*}

Figure \ref{fig:rho-dist} contains a plot of the distribution of the
density in Simulation 4 at $t=0$ and at $t=0.04$.  The fluid enters at $x=0$ and, as
it is decelerated by the cosmic ray pressure, the density
inhomogeneities are amplified as the regions of high density are
compressed.  Vorticity is generated as the regions of low density are
decelerated to a greater degree than regions of high density.  This
creates shocks at the front of the high density regions.

Ultimately, by the time the fluid has reached $x=1$, the flow has
become quite disordered and appears turbulent.  We will only focus on
the magnetic field amplification as this is the topic of this paper.  A
general study of turbulence driven in this manner would be interesting,
but beyond the scope of this work.

It is clear from Figure \ref{fig:res-study} that the magnetic field is
strongly amplified by the vorticity generated in the fluid by the action
of the cosmic ray pressure.  At the highest resolution amplification of
a factor of around 20 is achieved.  This gives an average magnetic field
amplification in the pre-shock region, which is the diffusion scale of
the highest energy cosmic rays, of around 10.  Recalling that this is a
lower limit (see Sect.\ \ref{sec:res-study}) it is clear that this
instability is certainly capable of amplifying the ISM magnetic field up
to levels which match observations: taking an initial field of
5\,$\mu$G and amplifying this to around 50\,$\mu$G through this
instability, and then passing it through the (strong) supernova
blastwave will amplify it to around 200\,$\mu$G which is in the range
required to explain the observed morphology of the X-ray synchrotron
emission \citep{2003ApJ...584..758V, 2003A&A...412L..11B, 2004ApJ...602..257B,
2005ApJ...621..793B, 2005A&A...433..229V, 2006AdSpR..37.1902B,
2006A&A...453..387P, 2012A&ARv..20...49V,Uchiyama:2007ly,2004A&A...416..595Y}.

\subsection{Dependence on RMS density fluctuations}

\begin{figure}
\includegraphics[width=9cm]{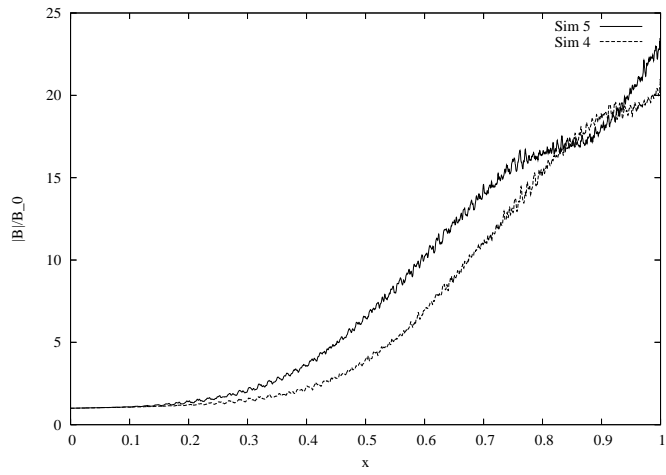}
\caption{ \label{fig:delta-rho-dep} Plots of the average magnetic field
	amplification as a function of $x$ for Simulations 4 and 5.}
\end{figure}

Using Simulations 4 and 5 we briefly investigate the influence of
varying $(\delta \rho)_{\rm rms}$, increasing it from $0.2$ to $0.44$.
Figure \ref{fig:delta-rho-dep} contains plots of the magnetic field
amplification as a function of $x$ for each simulation.  In the early
stages of the development of the instability (i.e.\ for small $x$) the
magnetic field is amplified more rapidly for higher $(\delta rho)_{\rm
rms}$, as might be expected.  However, as the fluid progresses towards
$x=1$ the magnetic field becomes amplified to approximately the same
extent in each case.  This can be understood through noting that once the 
nonlinear effects become important (i.e.\ once the density inhomogeneities are 
significantly amplified), the overall level of turbulence and vorticity
induced in the flow are approximately equal, thus leading to a similar
level of magnetic field amplification.

Of course, one does expect that as $(\delta \rho)_{\rm rms}$ increases
the field amplification will also increase.  It appears, though, that
nonlinear effects give rise to a relatively weak dependence of field
amplification on $(\delta \rho)_{\rm rms}$, at least if equations
\ref{eqn:lambda-limit} and \ref{eqn:b_limit} are satisfied.

\subsection{Dependence on initial magnetic field strength}

\begin{figure}
\includegraphics[width=9cm]{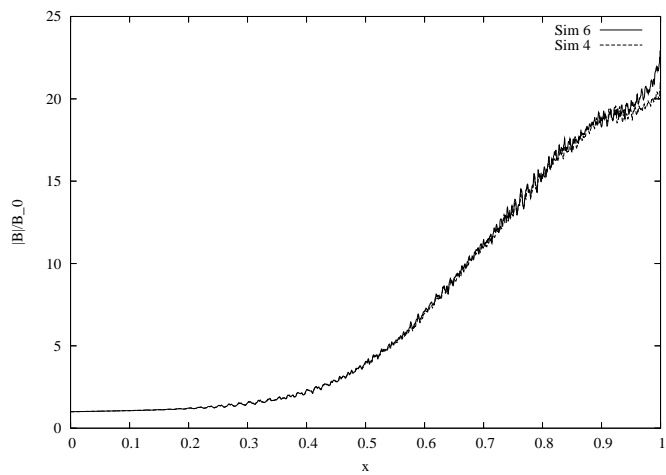}
\caption{ \label{fig:rms-dep} As Figure \ref{fig:delta-rho-dep} but for
	Simulations 4 and 6.}
\end{figure}

Figure \ref{fig:rms-dep} contains plots of the magnetic field
amplification as a function of $x$ for Simulations 4 and 6.  There is
virtually no dependence of the magnetic field amplification on the
initial field strength.  Again, this is dependent on equations 
\ref{eqn:lambda-limit} and \ref{eqn:b_limit} being satisfied.  However,
we can conclude that for conditions appropriate to supernova
blastwaves, the specific value of the magnetic field in the ISM is
not important in determining the final magnetic field amplification, at
least at the resolutions investigated here.

The energy density in the amplified magnetic field at $x=1$ is about 3\%
of that available in the velocity variations, $\delta u$, according to 
equation \ref{eqn:e_f}.  One would expect then that the back-reaction of the 
magnetic field on the turbulent flow would be negligible and thus the specific 
magnetic field strength should not be important when determining the final
amplification value.

\section{Conclusions}

We have demonstrated using a simple, but physically motivated and not unrealistic, model that significant magnetic field amplification can occur in cosmic-ray shock precursors driven by the differential acceleration of density inhomogeneities and the resultant turbulent vorticity field.  A more detailed model of the diffusion of cosmic rays into the precursor would probably lead to even higher amplification because the effective diffusion would be smaller in the amplified dense knots leading to larger pressure gradients.  

A noteworthy feature of the mechanism proposed here is that it operates up to what is essentially the largest scale available for any cosmic-ray driven process, the length scale of the precursor itself.  Clearly no cosmic-ray driven process can operate further ahead of the shock than the length scale determined by the diffusion of the most energetic particles.  It also has the great advantage of operating on scales large compared to the gyro-radius of the driving particles (at least in non-relativistic shocks where even in the Bohm limit the precursor length scale exceeds the gyro-radius by a factor of $c/U$, the ratio of the speed of light to the shock speed and typically several hundred to a thousand for SNR shocks) so that the amplified field can reduce the gyro-radius and trigger a boot-strap process.   Of course other processes are possible and may add to the field amplification, particularly on smaller scales where plasma kinetic effects such as those considered by Bell can operate.  The point we want to make here is simply that independent of all the detailed plasma physics, as long as there is a cosmic-ray precursor with a significant associated pressure gradient, and as long as the inflowing medium is clumpy, a well-stirred and significantly amplified magnetic field can easily be created in the precursor on the scales required.

\section*{Acknowledgements}
The authors wish to acknowledge the SFI/HEA Irish Centre for High-End Computing (ICHEC) for
the provision of computational facilities and support.

\printbibliography

\label{lastpage}

\end{document}